# Charge fluctuations and electron-phonon coupling in organic charge-transfer salts with neutral-ionic and Peierls transitions

A. Girlando[a], A. Painelli[a], S.A. Bewick[b] and Z.G. Soos[b]

[a] Dip. Chimica G.I.A.F., Univ. di Parma, I-43100 Parma, and INSTM-UdR Parma, Italy

[b] Department of Chemistry, Princeton University, Princeton NJ 08544, USA

## Abstract

The first-order transition of the charge-transfer complex TTF-CA (tetrathiafulvalene-chloranil) is both a neutral-ionic and a Peierls transition. In related organic charge transfer complexes, cooling at ambient pressure increases the ionicity $\rho$ in strikingly different ways, and is sometimes accompanied by a dielectric peak, that we relate to lattice stiffness, to structural and energetic disorder, and to the softening of the Peierls mode in the far-IR. The position operator $P$ for systems with periodic boundary conditions makes possible a systematic treatment of electron-phonon interactions in extended donor-acceptor stacks in terms of correlated Peierls-Hubbard models. The IR intensity of the Peierls mode peaks at the Peierls transition at small $\rho < 1/2$ in soft lattices, where the dielectric constant also has a large peak. In dimerized stacks, the IR intensity of totally symmetric, Raman active, molecular vibrations is related to charge fluctuations that modulate site energies. Combination bands of molecular and Peierls modes are identified in regular TTF-CA stacks above $T_c$. Energetic disorder can suppress the Peierls transition and rationalize a continuous crossover from small to large $\rho$. The TTF-CA scenario of a neutral-regular to ionic-dimerized transition must be broadened considerably in view of charge transfer salts that dimerize on the neutral side, that become ionic without a structural change, or that show vibrational evidence for dimerization at constant $\rho < 1$.

## 1. Introduction

Crystals of planar π-electron donors (D) and acceptors (A) typically contain face-to-face stacks, DADA, with significant π-overlap indicated by less than Van der Waals separation between molecular planes.[1] Such stacks are called *mixed* to distinguish them from *segregated* stacks of cation ($D^+$) or anion ($A^-$) radicals in other systems. Charge-transfer (CT) excitations from D to A or between radical ions are polarized along the stack and have interesting vibronic consequences[2-4] that also apply to conjugated polymers.[5,6] CT crystals of weak donors and acceptors have neutral ground states (gs) with partial ionicity $\rho < 0.1$. The strongest D and A form radical ion salts with $\rho \sim 1$ whose Madelung energy M exceeds the energy cost, $I - A$, to transfer an electron. Although rare, there are CT crystals with intermediate ionicity. The 1:1 salt D = TTF (tetrathiafulvalene) and A = CA (chloranil) studied by Torrance and coworkers[7] belongs to an even rarer class in which $\rho$ changes appreciably with temperature or pressure. On cooling below $T_c = 81$ K, TTF-CA has a valence or neutral-ionic transition (NIT) from a largely neutral ($\rho \sim 0.3$) to a largely ionic ($\rho \sim 0.6$) gs. Moreover, since the neutral stack has regular spacing while the ionic stack is dimerized, TTF-CA has a concomitant Peierls transition at the NIT.[8] Subsequent investigations[9-11] have produced a detailed and still evolving picture of the NIT in TTF-CA. It is remarkable to have electronic energies in balance on the scale of thermal energies.

Tokura, Horiuchi and coworkers[12-17] have recently increased the number of NIT systems by modifying either TTF or CA. They have also drawn attention to a large peak in the dielectric constant as a function of temperature $T$ or pressure $p$ and discuss their results in terms of quantum phase transitions from neutral-regular to ionic-dimerized lattices.[17] We summarize in Section 2 some salient features of these and other CT salts in order to underline their variability. A major purpose of this paper is to initiate the modeling of different types of crossovers from neutral to ionic gs. We will also develop the vibrational and electronic consequences of charge fluctuations in these systems. NITs with continuous $\rho(T)$ or $\rho(p)$ have already been recognized and modeled.[18-22] Indeed, long-range Coulomb interactions[23] or coupling to molecular vibrations[21] is needed for discontinuous $\rho$. For reasons given in Section 3, we expect soft lattices to dimerize on the neutral side, with $\rho < 1/2$ throughout, and the dielectric peak to mark the Peierls transition $T_P$. We consider in Section 4 how energetic disorder within single crystals can produce a continuous crossover from neutral to ionic in stacks that are regular according to X-ray data. There are also CT crystals with constant intermediate ionicity, independent of $T$. These observations clearly point to richer behavior than the TTF-CA paradigm of neutral-regular or ionic-dimerized.

Very schematically, Fig. 1 depicts mixed regular and dimerized stacks of centrosymmetric donors and acceptors. The inversion center at each D or A in the regular stack is retained in many structures in which molecules are tilted relative to the stack. Hence the gs of a regular stack is nonpolar, with equal transfer of charge from D to A in either direction on going between the neutral and ionic limits. The Peierls mode is the optical phonon that corresponds to motion of the D and A sublattices. By contrast, the gs of the dimerized stack is polar, the gs dipole reverses if the opposite dimerization is chosen, and CT crystals are potentially ferroelectric if all stacks dimerize the same way. Two choices for dimerization and one valence electron per site are characteristic of Peierls systems. Transition metal oxides with similar valence transitions and ferroelectricity have been discussed[24,25] using identical Hubbard models with broader bands than in organic CT salts.

The electronic problem in Hubbard models deals with the HOMO of D, the LUMO of A, and electron transfer $-t(1 \pm \delta)$ along the stack, with $\delta = 0$ in the regular stack. We have narrow bands ($4t \sim 0.5$-$1$ eV) and strong electron-electron correlations that exclude $D^{2+}$ or $A^{2-}$ sites. Different site energies $-\Gamma$ for D and $\Gamma$ for A is the basic modification of Hubbard models.[26] The gs ionicity depends on $\Gamma$, with large positive $\Gamma$ for the neutral lattice, large negative $\Gamma$ for the ionic lattice and a crossover around $\Gamma \sim 0$. Linear coupling of electronic degrees of freedom to harmonic phonons that modulate CT integrals or site energies is important to understand the phase transitions of the system and their rich spectroscopy. Even at the schematic level of Fig. 1, we have a model with many applications, successes and open questions.

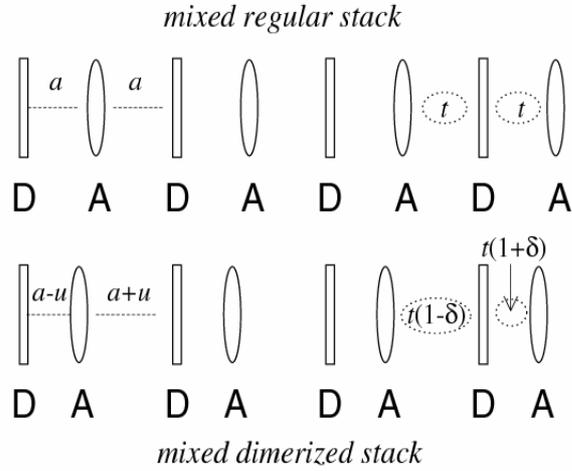

Fig. 1. Schematic representation of mixed regular and dimerized stacks, with spacing $a$, $a \pm u$ and transfer integrals $t$, $t(1 \pm \delta)$.

We note that although transitions are induced by temperature or pressure, their modeling is based on gs considerations. CT salts are insulators with large energy gaps compared to $k_BT$. In contrast to the standard metal-insulator Peierls transition, excited-state populations are negligible and we have quantum transitions between different ground states. Physically, since volume changes modulate $\Gamma$ and $t$, the gs evolves in response to an external parameter and sometimes does so discontinuously. Our modeling is not quantitative, however, since the precise relation between $\Gamma$ or $t$ and $T$ or $p$ is not known.

Vibrational spectroscopy provides accurate information on the gs properties of the material, as noted in Section 3. Since the frequencies of molecular modes are usually different in the molecule (either A or D) and the corresponding molecular ion (either $A^-$ or $D^+$), the ionicity can be determined from the vibrational frequencies in the solid. Rice first pointed out[2] the spectroscopic consequences of electron-molecular-vibration (e-mv) coupling in segregated stacks, which Pecile and coworkers extended to mixed stacks.[27] Electron-phonon (e-p) coupling in polyacetylene, $(CH)_x$, involves C-C stretches that, in the one-dimensional picture in Fig. 1, correspond to the molecular displacements modulating the dimerization amplitude. Simultaneous treatment of e-mv and e-p remains a challenging task. Fairly complete and sometimes quantitative modeling of vibrational spectra in CT crystals has been achieved using symmetry arguments, reference systems, and oligomers. In particular, the ionicity and linear e-mv coupling constants $g_\alpha$ have been determined for prototypical systems.[3,4] Important non-adiabatic effects are expected near a discontinuous transition, but will not be discussed here.[28]

Judicious use of models based on CT dimers or trimers accounts surprisingly well for many aspects of extended stacks in CT crystals.[3,4] There are basic limitations, however, related to the polarity of the gs in insulators with periodic boundary conditions. The problem is addressed and resolved in the modern theory of polarization.[30,31] In Section 3 we apply the quantum mechanical position operator[31] $P$ in periodic systems to CT salts.[32] Partial derivatives of $P$ yield the dielectric constant and the IR intensities of both the Peierls and ts molecular modes.[32,33] Thus $P$ makes possible a broader and more consistent discussion of organic CT salts with neutral-ionic and Peierls transitions. In Section 4 we comment on the different kinds of disorder in single crystals containing polar D or A and introduce a simple treatment of energetic disorder. The Discussion touches on current modeling of CT salts and on the implications of TTF-CA combination bands of the Peierls mode and ts molecular vibrations.

## 2. CT salts with partial or variable ionicity

McConnell *et al.* accounted[34] for the sharp separation, at that time, of organic CT salts into neutral diamagnetic and ionic paramagnetic systems by a simple mean-field argument that gives $\rho = 0$ or $1$, but not in between. In terms of theory, finite transfer integrals $t = - \langle D^+ A^- | H | DA \rangle$ and electron-electron correlations lead to intermediate $\rho$ in Hubbard-type models.[26] A harmonic lattice and linear electron-phonon (e-p) coupling in such models can be used to describe the Peierls instability, as shown by Su, Schrieffer and Heeger for polyacetylene[35] and by Rice and Mele for an AB polymer[36] that can be read as an uncorrelated model of CT salts. Several conclusions follow unequivocally. First, neutral stacks are expected to be regular. Second, there is a Peierls transition to a dimerized stack as $\rho$ increases, with continuous or discontinuous $\rho$ depending on model parameters. Third, ionic stacks are dimerized, although the amplitude may be very small, since $\rho = 1$ is a spin-1/2 Heisenberg antiferromagnetic chain[34] and hence subject to a spin-Peierls instability.[37] Finally, the NIT of the regular lattice is an inaccessible metallic point.[18] Attention has principally focused on the first-order transition of TTF-CA with a $\Delta\rho$ jump of ~0.2 at $T_c = 81$ K.

Table 1 summarizes some observations about CT salts with partial ionicity $\rho(T)$ at ambient pressure; donor and acceptor acronyms are listed at the bottom. Although our subsequent modeling of stacks is based on point molecules, or sites, it is important to recognize their internal degrees of freedom. Vibrational spectroscopy exploits the coupling of molecular modes to charge fluctuations along the stack. Molecules with $D_{2h}$ symmetry (aside from methyl groups) represent the majority of donors and acceptors used in ion-radical or CT salts, including organic conductors and superconductors, most likely through a combination of convenience and accident. This is not always the case in Table 1. The acceptors A = $QBr_nCl_{4-n}$ range from CA (n = 0) to bromanil (BA, n = 4);[15] the two n = 2 isomers are designated 2,5, which is centrosymmetric, and 2,6, which is not. The n = 1 and 3 isomers are polar, although only weakly so on chemical grounds. The acceptor $QCl_3H$ is more polar and forms crystalline alloys[13] $TTF(CA)_{1-y}(QCl_3H)_y$ over a wide range of y that also have large dielectric peaks around 40 K. The donors TTF, DMTTF, and TMPD have $D_{2h}$ symmetry (aside from methyl groups), as does TCNQ and the even more potent acceptor $TCNQF_4$. The ionicity of these D or A is related to changes of bond lengths or vibrational frequencies. $M_2P$ and other phenazines, by contrast, illustrate planar cation radicals and nonplanar neutral molecules.[45] $M_2P$-TCNQ is dimerized[46] at 300 K and has constant[42] $\rho \sim 0.5$, while $M_2P$-$TCNQF_4$ is regular and ionic.[41,42] The polar donor ClMPD has an unusual $\rho(T)$ crossover[40] in ClMPD-DMDCNQI that we suggest below to be due to disordered dipoles. Polar D and A are often disordered in crystals, with equal population of two orientations. That is the case for the $QBr_nCl_{4-n}$ series in ref. 15, for ClMPD in ref. 40, and for several D in segregated stacks.[47] The crystals are nevertheless centrosymmetric based on average D or A. We comment in the Discussion on the constant-$\rho$ systems TTF-BA[38], TMPD-CA[43] and TMPD-TCNQ.[44]

Table 1. CT salts with partial or variable ionicity $\rho$.

| CT crystal[a] | $\rho(T)$ range | $\rho(T)$ shape | $\kappa$ max | T (K), $\kappa$ peak | IR side bands | T(K), IR vibronic | Peierls, X-ray | Ref. |
|---|---|---|---|---|---|---|---|---|
| TTF-CA | 0.2 – 0.7 | 0.2 jump | 600 | 81 | Yes | 81 | Yes, 81 | 8-10 |
| TTF-BA | ~ 1.0 | Constant | – | – | _ | 50 | – | 38 |
| TTF-QCl$_3$H | 0.2 – 0.3 | Contin. | 50 | No peak | _ | 100 | No | 13 |
| TTF-QBrCl$_3$ | 0.2 – 0.7 | 0.1 jump? | 800 | 66 | Yes | 66 | Yes | 14,16 |
| TTF-2,5QCl$_2$H$_2$ | 0.2 - 0.3 | Contin. | - | - | Yes | No, to 15 | - | 39 |
| DMTTF-CA | 0.3 – 0.4 | S-shape | 200 | 65 | Yes | 68 | Yes, 68 | 15 |
| DMTTF-2,5 QBr$_2$Cl$_2$ | 0.2 – 0.4 | S-shape | 200 | 30 | Yes | 49 | _ | 15 |
| DMTTF-2,6 QBr$_2$Cl$_2$ | 0.2 – 0.3 | Contin. | 190 | No peak | Yes | 25 ? | _ | 15 |
| DMTTF-BA | 0.2 – 0.3 | Contin. | 80 | No peak | Yes | No, to 4 | _ | 15 |
| ClMPD-DMDCNQI | 0.3 – 0.6 | S-shape | – | – | _ | 320 | No | 40 |
| M$_2$P-TCNQ | 0.5 | Constant | – | – | _ | 300 | >300 | 41,42 |
| M$_2$P-TCNQF$_4$ | >0.9 | Constant | – | – | _ | 120 | – | 41,42 |
| TMPD-CA | 0.6 | Constant | – | – | _ | 300 | No | 43 |
| TMPD-TCNQ | 0.9 | Constant | – | – | _ | 300 | No | 44 |

[a] CA = chloranil, QCl$_4$; BA = bromanil, QBr$_4$; TTF = tetrathiafulvalene, DMTTF = dimethylTTF; TMPD = tetramethyl-*p*-phenylenediamine; M$_2$P = dimethylphenazine; TCNQ = tetracyanoquinodimethane; ClMPD = chloromethyl-*p*-phenylenediamine; DMDCNQI = dimethyldicyanoquinodiimine.

The ionicity $\rho(T)$ in the second and third column of Table 1 is primarily based on vibrational spectroscopy. The indicated range is between $\rho(300K)$ and $\rho(low)$, the lowest reported temperature. We distinguish among cooperative $\rho(T)$ variations, either discontinuous or S-shaped, and slowly varying or constant $\rho$. The dielectric data and its temperature dependence is limited to the indicated salts and extends to ~ 4K. While DMTTF-BA has no peak on cooling, it has a $\kappa$ peak under a few kbar of pressure.[17] The "IR side bands" column refers to IR bands seen in correspondence to, but not in coincidence with, ts modes before the Peierls transition. As discussed in Section 3, side bands are combinations of the Peierls and ts modes that appear in regular stacks via coupling to $\pi$-electrons. The adjacent column indicates the temperature at which vibronic bands, coincident in frequency with the Raman, appear in IR spectra. The activation of Raman ts modes in IR indicates the loss of the inversion site symmetry. In some cases the Peierls distortion has been confirmed by X-ray data, as indicated in the rightmost column, and when available the temperature of the structural transition is listed.

Table 1 illustrates the range of behavior in CT salts with neutral-ionic transitions or partial ionicity. Although far from complete either in terms of systems or measurements, these data clearly indicate the need to generalize the TTF-CA paradigm. As already mentioned, one distinction is lattice stiffness that allows a Peierls instability in soft lattices at small $\rho$, below the crossover to an ionic gs. The order parameter for the second-order transition is the dimerization amplitude.[20,21] Another distinction involves structural disorder in single crystals of polar D or A. Disorder can in principle suppress the Peierls instability and stabilize ionic lattices with regular stacks based on X-ray structural determination.

## 3. Charge fluctuations and IR spectra

Quadratic coupling between electronic and vibrational degrees of freedom accounts for different vibrational frequencies in different electronic states, and affects, in variable amount, all vibrational modes.[3] It is mainly responsible for a renormalization of the reference vibrational frequencies, and its role in t-PA has been extensively discussed.[5] In CT salts, it provides the basis for an accurate determination of $\rho$ based on the comparison of the frequencies of non-totally symmetric modes with those of reference compounds with known ionicity. For example,[8] the $b_{1u}\nu_{10}$ mode of CA involving the asymmetric C=O stretch has higher frequency by 160 cm$^{-1}$ in the neutral ($\rho = 0$) crystal than in an alkali-CA salt ($\rho = 1$). Several other modes of D or A have significant shifts of over 25 cm$^{-1}$ between systems whose $\rho = 0$ or 1 is fixed by the chemistry. The frequency in a CT crystal yields the ionicity on assuming a linear dependence on $\rho$. The procedure's consistency can be checked when there are several shifted modes of D and A.

The most interesting features in vibrational spectra of materials with delocalized electrons are governed by linear electron-phonon coupling[2-5] that, in CT salts, involves both molecular vibrations and lattice phonons. As Rice noticed for segregated stacks, upon ionization D and A molecules relax along vibrational coordinates that are ts with respect to the molecular point group. The resulting linear dependence of the on-site energy on ts molecular coordinates, called linear electron-molecular vibration (e-mv) coupling, corresponds to Holstein coupling[48] and the relevant relaxation energy is the small-polaron binding energy $\varepsilon_{sp}$ as in polaron theory. Relaxation on harmonic surfaces appears in other contexts. It is the Franck-Condon energy for vertical excitations in spectroscopy or the reorganization energy $\lambda$ in Marcus theory of electron transfer, to name just two.

Holstein coupling is responsible for several interesting features in vibrational spectra of CT salts. First of all, the frequencies of ts modes are significantly softened with respect to the frequencies expected for a molecule with the same ionicity as in the crystal. For this reason only the frequencies of non-ts modes can be used to estimate $\rho$ as described above.[3] Even more interesting are the effects of Holstein coupling on IR intensities. Totally symmetric molecular modes are Raman active, IR forbidden in isolated molecules with inversion symmetry or in stacks where the molecules reside on an inversion center. In dimerized stacks, or more generally in stacks where molecules do not reside on inversion centers, on-site ts vibrations induce asymmetric charge flow along the chain, then inducing a dipole along the stack.[3] Thus ts vibrations in dimerized stacks appear both in IR and Raman spectra at the same frequency, as noted in Table 1. Since charge fluxes occur along the chain, the coupled modes are polarized along the chain and can be very intense by borrowing from the CT transition. Hence Raman and IR spectra distinguish sharply between regular and dimerized stacks of centrosymmetric molecules.

Peierls coupling accounts for the linear dependence of the CT integrals on the intermolecular distance. In a perfect DA crystal it primarily involves the k=0 optical lattice phonon that modulates the dimerization amplitude. The same coordinate, called the effective conjugation coordinate,[6] is the Peierls mode of *trans*-PA and is responsible for the intriguing spectroscopic behavior of this prototypical polymer[5] as well as of other conjugated polymers.[6] In particular, the extended conjugation coordinate is distributed over the three ts modes of the *trans*-PA lattice in the mid-IR region. The Peierls coupling is then responsible for the large softening of the corresponding frequencies with increasing the conjugation length, as evidenced by the large dispersion of the relevant Raman frequencies with the excitation line. The same three modes acquire a huge IR intensity in doped and photoexcited samples where solitons appear. The Peierls mode of CT crystals is IR allowed and is expected in the far-IR spectral region, since lattice force constants are much smaller than in polymers, and effective masses are much larger. Much as in *trans*-PA, the Peierls mode is expected to soften in CT salts due to e-p coupling, but at variance with *trans*-PA, due to the reduced symmetry of the stack, it acquires huge IR intensity at the dimerization phase transition.[32]

The simplest Hamiltonian for a CT stack with periodic boundary conditions is an N-site Peierls-Hubbard model with alternating site energies $\Gamma$.[18] Its gs properties are given by

$$H = \sum_p (-1)^p \Gamma n_p - \sum_{p,\sigma} [1 - (-1)^p \delta](c_{p\sigma}^+ c_{p+1\sigma} + h.c.) + N \frac{\delta^2}{2\varepsilon_d} \qquad (1)$$

Here $n_p$ is the number operator and $c_{p\sigma}^+$ creates an electron with spin $\sigma$ at site $p$, with D at odd $p$ and A at even $p$. The transfer integral of the regular ($\delta = 0$) stack is the unit of energy ($t = 1$). Peierls coupling adds a harmonic potential $\delta^2/2\varepsilon_d$ per site for a lattice with stiffness $1/\varepsilon_d$. This corresponds to the standard model,[35] with $t$ linearly depending on the site displacements $u_i$ : $t_i = t_0 - \alpha(u_i - u_{i+1})$, where $\alpha = (k\varepsilon_d/2)^{1/2}$ and $k$ is the bare force constant. The number of electrons equals the number of sites: the $\rho = 0$ stack has two electrons paired in the HOMO of D; the $\rho = 1$ stack has one electron in the HOMO of D and one in the LUMO of A. We exclude high-energy $D^{2+}$ and $A^{2-}$ sites on physical grounds. The electronic system is then highly correlated.

The exact gs, $\psi(\delta,\Gamma)$, is a singlet ($S = 0$) with energy $\varepsilon(\delta,\Gamma)$ per site. The ionicity is $\rho = 1 + \partial \varepsilon/\partial \Gamma = \langle n_{2p} \rangle$ at A sites. The total gs energy per site is

$$\varepsilon_T = \varepsilon(\delta,\Gamma) + \delta^2/2\varepsilon_d \qquad (2)$$

The NIT of the $\delta = 0$ lattice is a metallic point at $\Gamma_c = -0.666$ where $-(\partial^2\varepsilon/\partial\delta^2)_0$ diverges,[18] while the Peierls transition occurs at $\Gamma_P(\varepsilon_d) > \Gamma_c$ where $(\partial^2\varepsilon/\partial\delta^2)_0 + 1/\varepsilon_d = 0$. The equilibrium dimerization $\delta$ at $\Gamma < \Gamma_P$ is found by minimizing $\varepsilon_T$ with respect to $\delta$. The model (1) requires a minimum stiffness of $1/\varepsilon_d \geq 1/\sqrt{2}$ that leads to $\delta = \pm 1$ at $\Gamma = 0$. Exact solutions up to $N \sim 20$ yield accurate extrapolations[18] to the infinite stack, such as $\rho(0,\Gamma_c) = 0.684$. The crossover is asymmetric with respect to $\rho = 0$ and 1 because CT is quite different in neutral stacks with paired spins on D and stacks of radical ions with a singlet gs. The asymmetry is entirely due to correlations: direct solution of (1) for an infinite stack without excluding $D^{2+}$ or $A^{2-}$ yields a symmetric crossover at $\Gamma = 0$ where the two sublattices are interchanged.

The NIT and Peierls instability are general features: only their location depends on the chosen model. By contrast, continuous or discontinuous variations of $\rho(\delta,\Gamma)$ depend on the model, and (1) has continuous $\rho$. The energy difference, $2\Gamma = E(D^+A^-) - E(DA)$, clearly involves Coulomb interactions in addition to $I - A$ for transferring an electron. Long-range Coulomb interactions can be added to (1) and lead to discontinuous $\rho$ for narrow enough bands.[21,23,26] Discontinuous $\rho$ is expected and found in mean-field treatments when the crystal electrostatic (Madelung) energy exceeds a critical value.[21,26] When D and $D^+$ or A and $A^-$ have the same point group symmetry, changes in bond lengths and angles between the molecule and ion correspond to shifts $\Delta Q$ along ts normal modes. The site energy $\Gamma(\rho)$ then varies with ionicity and sufficiently strong Holstein coupling also yields discontinuous $\rho$. The coupling of the Peierls mode to the electronic systems is explicit in (1), while Coulomb interactions or Holstein coupling are extensions.

The Peierls mode of the regular stack is IR active. Its intensity comes mainly from polarization of the electronic system. In units of e$a$, where $a$ is the DA separation of a regular stack, the electronic polarization per site of (1) is[31,32]

$$P(\delta,\Gamma) \quad = \quad \frac{1}{2\pi} \operatorname{Im} \ln \langle \psi(\delta,\Gamma) | \exp(\frac{2\pi i}{N} \sum_p p q_p) | \psi(\delta,\Gamma) \rangle \qquad (3)$$

The charge operator is $q_{2p-1} = 2 - n_{2p-1}$ at D sites and $q_{2p} = -n_{2p}$ at A sites. The polarization of a periodic system is given by a phase, as in Berry phases, and the usual dipole operator appears in the exponent. The gs expectation value is real by symmetry at $\delta = 0$. For finite $N$ and exact $\psi(\delta,\Gamma)$, we evaluate the expectation value in (3) as $Z_x(\delta,\Gamma) + iZ_y(\delta,\Gamma)$ and obtain $2\pi P(\delta,\Gamma) =$

arctan($Z_y$/$Z_x$). As emphasized by Resta,[31] the derivatives of $P$ rather than $P$ itself are related to physical observables. In the present case, the Peierls mode drives strong oscillations of charge along the stack as partial double bonds are formed with one neighbor for $\delta > 0$ and the other for $\delta < 0$. The electronic IR intensity goes as $(\partial P/\partial \delta)^2$ and far exceeds[32] the conventional contribution due to the motion of frozen charges $\pm \rho e$.

The charge flux, $\partial P/\partial \delta$, is a function of $\delta$ and $\Gamma$. It is small for large positive or negative $\Gamma$, where there is little CT mixing, or for strong alternation $\delta > 0.5$, where there is little delocalization. The flux for the equilibrium dimerization gives the IR intensity of the Peierls mode. Starting with large $\Gamma$ and $\delta = 0$ on the N side, $(\partial P/\partial \delta)^2$ increases with decreasing $\Gamma$ and would diverge at $\Gamma_c$ were it not for the Peierls transition at $\Gamma_P(\varepsilon_d) > \Gamma_c$. The intensity peaks at $\Gamma_P(\varepsilon_d)$ because dimerization more than offsets decreasing $\Gamma$. Moreover, the peak intensity increases with the lattice stiffness since the instability is then closer to the NIT where charges are most mobile. These expectations are born out in Fig. 2 by direct calculations for N = 10 and 12 in a stiff lattice with $\varepsilon_d = 0.28$, $\Gamma_P = 0.0$ that approximates TTF-CA and a softer one with $\varepsilon_d = 0.64$, $\Gamma_P = 0.8$. The peak intensity is an order of magnitude greater in the stiff chain. The behavior of $\rho(\delta_{eq}, \Gamma)$ is similar in both cases, while the maximum dimerization of the soft chain is almost four times larger. Both chains have vanishing $(\partial P/\partial \delta)^2$ near $\Gamma_c$, the NIT of the regular chain, that we interpret as a reversal of the charge flux. Dimerization increases $(1+\delta)$ and hence CT mixing, thereby increasing $\rho$ in a neutral stack and decreasing $\rho$ in an ionic stack.

Totally symmetric molecular vibrations are only Raman active in centrosymmetric regular lattices. They acquire IR intensity in dimerized stacks due to Holstein coupling, which modulates $\Gamma$ and therefore $\rho$. To model this effect quantitatively, we replace $\Gamma$ in the Hamiltonian (1) by

$$\Gamma \quad \rightarrow \quad \Gamma + \sqrt{\varepsilon_{sp}/N} Q \tag{4}$$

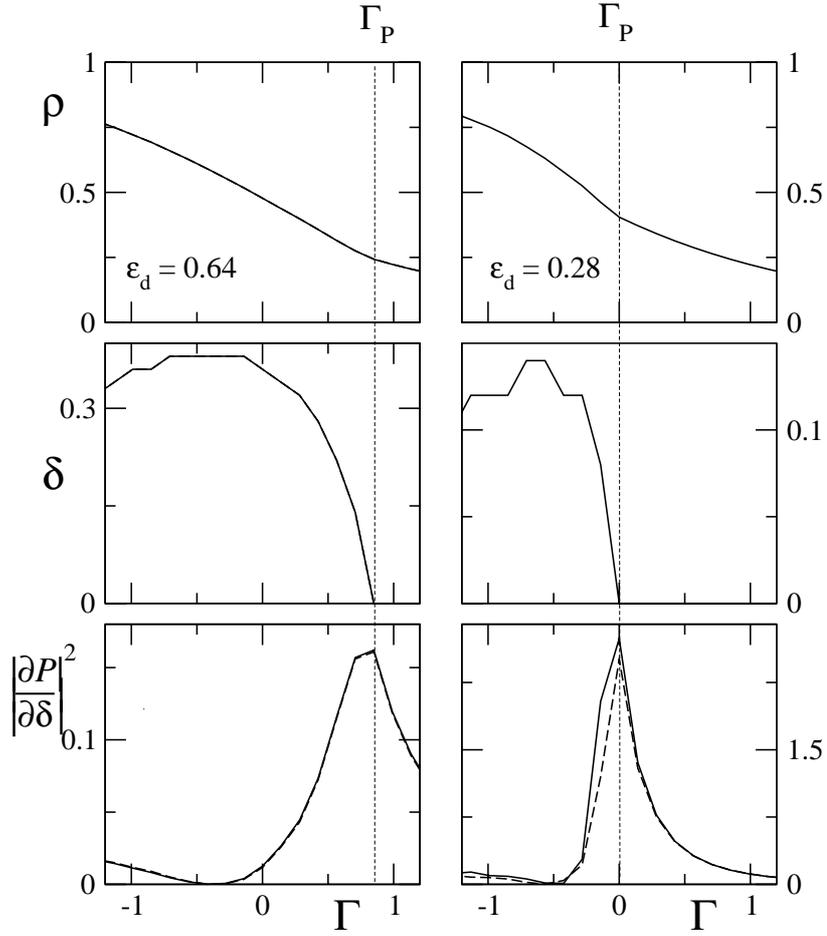

Fig.2. Ionicity, dimerization amplitude and IR-intensity of the Peierls mode vs. $\Gamma$ for two values of $\varepsilon_d$. Continuous and dashed line are for 12 and 10 sites, respectively, that coincide in most panels. The line across the panels locates the Peierls transition.

where $Q$ is the coordinate describing the in-phase (k=0) vibration of all A (or equivalently D) molecules along a ts molecular coordinate. The potential $\omega_M^2 Q^2 / 2$ is also added to $H$ in the harmonic approximation.[33] The gs geometry in the adiabatic approximation now requires energy minimization not only with respect to $\delta$, as above, but also with respect to $Q$. Holstein coordinates affect the on-site energy (cf 4, above), and, in turn, are affected by $\rho$, leading to a self-consistent problem that originates cooperativity.

Figure 3 shows representative $(\partial P/\partial \Gamma)^2$ results for the soft lattice with $\varepsilon_d = 0.64$ and $\varepsilon_{sp} = 0.28$ or 2.8. The IR intensity of course vanishes for $\Gamma > \Gamma_P$. As $\varepsilon_{sp}$ increases, $\rho(\delta_{eq}, \Gamma)$ develops a kink at $\Gamma$ ̣̣̣̣̣̣̣̣̣ ................................... stiffer lattice.[33] The IR intensity

asymmetr

larger $\varepsilon_{sp}$

rapidly th

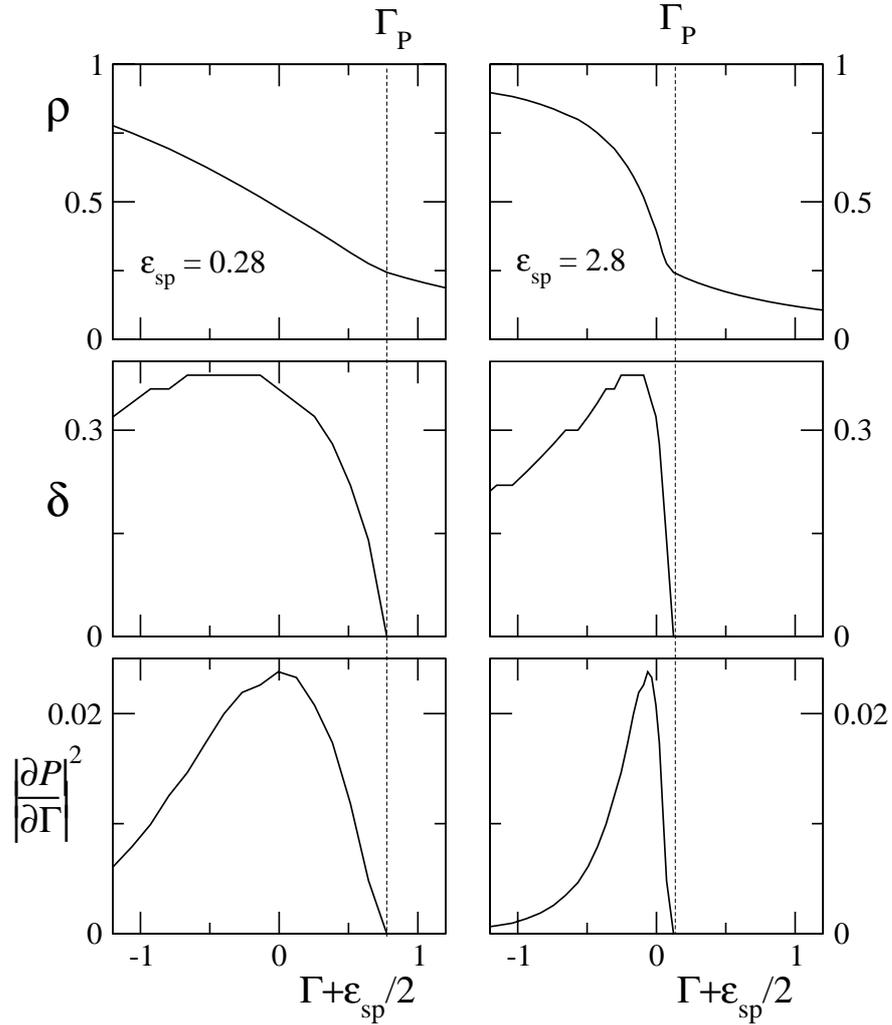

Fig. 3. Ionicity, dimerization amplitude and IR-intensity of the Holstein mode vs. $\Gamma$ for a 10 site ring with $\varepsilon_d = 0.64$ and two values of $\varepsilon_{sp}$. The line across panels locates the Peierls transition.

IR absorptions at precisely the same frequency as ts Raman bands are expected and found in dimerized stacks. More puzzling is the observation of side bands in several systems listed in Table 1. As mentioned in Section 2, several salts in the high-T (regular) phase show weak IR absorption bands at frequencies that are close to, but not coincident with, the frequencies of Raman peaks. These IR *side-bands* have been attributed to ionic domains in neutral crystals,[14,15] but a careful study of TTF-CA allowed us to identify the Peierls mode and to propose an alternative explanation.[49]

Although both lattice and molecular vibrations are harmonic by hypothesis, the gs potential energy surface becomes anharmonic, sometimes strongly so, when e-p coupling is taken into account.[50] Moreover, charge fluctuations provide a coupling between lattice and molecular modes that rationalizes combination bands whose modeling requires relaxing the harmonic approximation.[49] Qualitatively, the combination band at $\omega_+ = \omega_M + \omega_P$ is IR allowed and polarized along the stack. So is the hot band at $\omega_- = \omega_M - \omega_P$ whose intensity depends on $k_B T > \hbar\omega_P$ since a Peierls phonon is annihilated. Hence IR combination bands polarized along the stack are expected on either side of ts modes at $\omega_M$ in the Raman spectrum. For TTF-CA, three different Raman bands have been identified showing side-bands in the IR.[49] Figure 4 shows the difference between the corresponding $\omega_+$ and $\omega_-$ (left and right panels, respectively) and the relevant Raman frequency. The data indicate that all the combination bands originate from the same far-IR vibration whose frequency undergoes a pronounced softening with decreasing $T$. The solid line in both panels is the fit to $\omega_P = b\sqrt{T - T_P}$, the standard soft-mode expression, with $b = 4.6$ cm$^{-1}$ K$^{-1/2}$ and $T_P = 57$ K. We identify this mode as the Peierls mode, whose softening is consistent with a Peierls transition at 57 K that is subsumed by the first-order TTF-CA transition at $T_c = 81$ K. The soft-mode interpretation of Fig. 4 is natural in the present context and illustrates the evolution of TTF-CA modeling. We return to the alternative view of ionic domains in the Discussion. A soft-mode interpretation bears directly on systems in Table 1 with side bands and also indicates the need for improved treatment of vibrations.

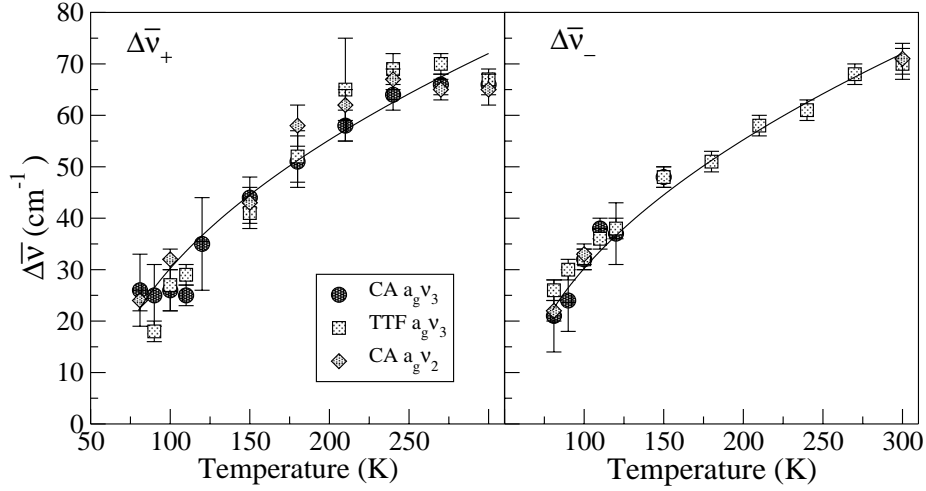

Fig. 4.  Temperature dependence of the frequency difference between IR bands polarized along the stack and the indicated fundamental Raman bands $\omega_M$ of CA or TTF, from Ref. 49. Data in the left and right panels refer to the combination bands $\omega_+ = \omega_M + \omega_P$ and $\omega_- = \omega_M - \omega_P$. The full line represents the fit to a standard expression for a soft mode (see text).

The static dielectric constant $\kappa$ is directly related to the polarizability $\alpha$ per unit cell. The model (1) describes CT processes but not the electronic excitations leading to $\kappa_\infty \sim 3$ in organic molecular solids. The $\kappa$ peaks in Table 1 are one to two orders of magnitude larger and $\alpha$ contains both electronic and vibrational contributions,[32]

$$\alpha \;\; = \;\; \alpha_{el} + 2\,\mu_{IR}^2/\Omega \qquad\qquad (5)$$

The IR intensity is related to $\varepsilon_d(\partial P/\partial\delta)^2$, evaluated at $\delta_{eq}$, while $\Omega = \partial^2\varepsilon_T/\partial\delta^2$ is the curvature of the gs potential surface in (2). The electronic part $\alpha_{el}$ is the linear response to a static electric field F. The interesting issue of combining a static electric field with periodic boundary conditions has recently been addressed[51,52] in metal oxides in terms of $P$. A related procedure[53] yields the exact static (hyper-)polarizabilities of the Peierls-Hubbard model (1), with $\alpha_{el} = (\partial P/\partial F)_0$. As also found in uncorrelated systems,[32] $\alpha_{el}(\delta_{eq},\Gamma)$ increases as $\Gamma_P$ is approached from the neutral side and then varies slowly. The IR contribution of the Peierls mode, by contrast, diverges at $\Gamma_P$ and damping has to be taken into account. Hence we associate a $\kappa$ peak with the Peierls transition, even though the electronic and vibrational contributions are of the same order. Estimates of the peak height are promising but rather rough for several reasons, including the treatment of damping.[53] In Table 1, the $\kappa$ peak marks Peierls transitions that coincide with the NIT when $\rho$ is discontinuous (TTF-CA)[14] or almost discontinuous (TTF-QBrCl$_3$, DMTTF-CA).[15,16]

## 4. Disorder in CT single crystals

Centrosymmetric D and A often crystallize in space groups that retain the inversion center at D and A. Symmetry then ensures a regular ($\delta = 0$) stack with equal $t$ between any neighbors. The symmetry argument is important because reliable computation of $t$ for such molecules is still challenging; $t$ depends strongly on both distance and the nodes of the frontier orbitals. We have so far considered neutral regular stacks with a Peierls transition to dimerized stacks and alternating $t = -(1 \pm \delta)$ as $\rho$ increases. Dimerization in the Peierls-Hubbard model (1) is strictly limited to charge fluctuations. The model is not appropriate for dimerization due to other causes. For example, the perchlorate anion stops rotating at the dimerization transition of TMPD-ClO$_4$, Wurster's blue perchlorate, and such phase transitions of organic stacks are not infrequent in systems with high-symmetry counterions.

The M$_2$P-TCNQ and M$_2$P-TCNQF$_4$ salts in Table 1 illustrate the consequences of gross structural differences between D and D$^+$ that are beyond the model (1). The M$_2$P-TCNQ structure shows[46] a mixed dimerized stack with unequal spacing in Fig. 1 and bent M$_2$P with a dihedral angle of 165°. There are obviously two $t$'s along the stack. Vibrational data yield $\rho \sim 0.5 \pm 0.1$ with coincident Raman and IR features as expected;[42] the ionicity and dimerization are also consistent with magnetic data such as a small singlet-triplet gap and triplet spin exciton spectra.[41] The stronger acceptor TCNQF$_4$ leads instead to an ionic ($\rho \sim 0.9$) complex containing planar M$_2$P$^+$ radical ions and mixed regular stacks at ambient temperature.[41] The stack dimerizes around 120 K, as indicated by both vibrational[42] and magnetic data,[41] but no low-temperature structure has been reported.

A different situation arises when guest molecules are incorporated in crystal lattices to yield doped crystals or alloys. TTF-CA forms alloys[13] with the weaker acceptor A = QCl$_3$H. Such DA$_{1-y}$A$_y$ systems with variable y clearly have energetic disorder along the stack that calls for variable $\Gamma_n$ in the model (1). More pervasive disorder appears when either D or A is not centrosymmetric and the crystal refines to mixed regular stacks based on the average structure. The DMTTF-QCl$_{4-n}$Br$_n$ crystals in Table 1 are disordered[15] with respect to Cl/Br. Similarly, ClMPD forms regular DA stacks[54] with A = DMDCNQI. Although the average structure has equal $\bar{i}$ by symmetry, the absence of an inversion center at either A or D leads to $t_1 \neq t_2$ for the two neighbors in the stack. It is then a quantitative issue to determine different $t$'s in a particular crystal and the importance of $t_1 \neq t_2$.

The dipoles $\mu$ of polar D or A are another source of disorder. Dipoles introduce long-range interactions in addition to electrostatic (Coulomb) interactions for charges $\rho > 0$. Dipoles due to Br/Cl are likely to be small. By contrast, ClMPD or ClMPD$^+$ is calculated[55] to have $\mu \sim$ 3D. The two orientations of ClMPD in the crystal give a random Ising model with $\pm \mu$ whose potentials and fields represent energetic disorder. Such disorder has been studied in connection with hole transport in molecularly doped polymers[56] (MDPs), which are amorphous systems of polar D embedded in polymers.

CT crystals with average structures of polar D or A introduce interesting new aspects that have nothing to do with imperfections or defects or overlap fluctuations, but are associated with "perfect" crystals. To the extent that face-to-face stacking yields a one-dimensional electronic system, we know that disorder leads to localized states. Since the gs is already localized on the neutral side of regular chains, disorder is not important qualitatively except near the NIT, where the metallic point is always suppressed. The Peierls instability at $\Gamma_P > \Gamma_c$ depends on the lattice stiffness $1/\varepsilon_d$ in the Peierls-Hubbard model of Section 3. Stiff lattices that require extensive delocalization for dimerization will be sensitive to disorder that limits delocalization, while the dimerization of soft lattices can only be suppressed by strong disorder.

To initiate the modeling of energetic disorder, we introduce a Gaussian distribution with variance $\sigma^2$ as in MDPs.[56] The A/D site energies in (1) are taken as $\pm(\Gamma+x_j)$ with $x_j$ chosen from the normalized distribution

$$f(x,\sigma) = (2\pi\sigma^2)^{-1/2}\exp(-x^2/2\sigma^2) \tag{6}$$

An N-site stack has random site energies with mean of zero. The gs energy per site is $\varepsilon(\delta,\Gamma,\sigma)$ and can be found for finite N via replicas. The gs energy $\varepsilon_r$ of replica r with N choices of $x_j$ is averaged over R replicas. The divergence of $-(\partial^2\varepsilon/\partial\delta^2)_0$ at $\Gamma_c$ is suppressed for any $\sigma > 0$, since disorder disrupts the delocalized gs at the NIT. Increasing $\sigma$ then makes $-(\partial^2\varepsilon/\partial\delta^2)_0$ less positive and eventually lowers it below $1/\varepsilon_d$ over the entire range of $\Gamma$. The *smallest* $\sigma_{min}$ that yields $\delta = 0$ at the minimum of $\varepsilon_T$ in (2) is the disorder needed to suppress the Peierls instability. As a first estimate, we suggest $\sigma_{min} = \Gamma(\varepsilon_d) - \Gamma_c$. Weak disorder suffices for stiff lattices whose Peierls transition requires considerable delocalization of the gs, while strong disorder is needed for soft lattices that dimerize readily. The $\varepsilon_d = 0.28$ and 0.64 lattices in Fig. 2 then require $\sigma > 0.7$ and 1.5, respectively, for a neutral-ionic crossover without dimerization.

As a simple example, we consider the Peierls-Hubbard model (1) with N = 4 and a small basis of seven singlets. The Jahn-Teller instability at $\delta = 0$, $\Gamma_c = -1/2$ has degenerate gs with $\varepsilon = -1/2$ per site, and the curvature $-(\partial^2\varepsilon/\partial\delta^2)_0$ is not defined at $\Gamma_c$ for $\sigma = 0$. An average over many replicas is needed in view of only four choices of $x_j$. As anticipated, $-(\partial^2\varepsilon/\partial\delta^2)_0$ decreases with increasing $\sigma$, and the instability is suppressed at $\sigma_{min} = 1.0$ and 2.0 for $\varepsilon_d = 0.13$ and 0.30, respectively. Since the instability of 4n rings becomes weaker with increasing n, smaller $\sigma_{min}$ will suppress the Peierls instability of the infinite stack, consistent with the $\sigma > 0.7$ estimate based on $\Gamma(\varepsilon_d) - \Gamma_c$. For $\sim 0.2$ eV, this gives $\sigma > 0.12$ eV and can be compared directly to lattices with randomly oriented dipoles $\mu$. The analysis[56] of hole transport in MDPs based on D's with ~1 D dipoles leads to $\sigma \sim 0.1$ eV. Hence the ~3 D dipole of ClMPD easily produces the required energetic disorder.

## 5. Discussion

The importance of charge fluctuations and electron-vibrational coupling in CT stacks has long been appreciated. Following the pioneering work of Rice on spectroscopic signatures of e-mv coupling,[2] a host of theoretical and experimental work[3,4] was devoted to the subject and vibrational spectroscopy is considered nowdays a useful tool to get reliable gs information in general. Several important microscopic parameters, including e-mv coupling constants, are routinely extracted from vibrational spectra.[57,58] However, the modeling of IR intensity of either mixed or regular stacks was unavoidably based on carefully chosen oligomers, where repeating dimeric, trimeric, etc. units were considered. Quantitative modeling of vibrational intensities in extended chains was not possible without the definition of the polarization $P$ in (3) for periodic systems. With $P$ in hand, its partial derivatives open the way for systematic treatment of vibrational intensities of the Peierls-Hubbard model (1) and of its more complicated versions. More discriminating modeling of the CT crystals in Table 1 can now be pursued.

In perfect lattices, ts molecular vibrations can only acquire IR intensity in the presence of dimerization. But local disorder breaks on-site inversion symmetry and rationalizes IR intensity for ts molecular vibrations in lattices that are regular according to X-rays. Careful comparison between vibrational and X-rays data is needed to define the gs structure precisely, particularly in disordered systems such as alloys or crystals of dipolar molecules. Explicit inclusion of disorder in CT crystals is clearly needed.

The implications of combination bands have to be considered. Intense IR bands associated with ts vibrations in dimerized stacks occur at frequencies coincident with those of intense Raman bands. The appearance of relatively weak bands in IR spectra of CT salts with regular stack can naturally be explained in terms of combination bands of a ts molecular vibration with a lattice phonon[49] as discussed for the TTF-CA spectra in Fig. 4. This interpretation is particularly suggestive in view of the large anharmonicity induced by e-ph coupling on Holstein and Peierls coordinates and is very useful, since quantitative information on the softening of the Peierls mode in the far-IR region can be obtained from a careful analysis of mid-IR spectra.

We comment briefly on the previous interpretation of IR side-bands as due to the presence of dimerized ionic domains in neutral regular stacks. [14,15] Following similar discussions for $(CH)_x$, organic CT salts have been analyzed in terms of solitons and neutral-ionic domain walls (NIDW). [11,15] There are spectroscopic clues for such ideas, for instance in electron spin resonance, [59] but quantitative treatment of NIDWs has proved to be difficult. The key observation [49] against ionic domains in TTF-CA is the CA mode at 1050 cm$^{-1}$ in the IR spectrum at 300 K. This is higher than the $a_g\nu_3$ mode at 980 cm$^{-1}$ in the Raman spectrum, and also higher than the $a_g\nu_3$ mode in either $\rho = 0$ or 1 systems. A combination band accounts for the frequency and such bands and must be considered seriously for other systems in Table 1 with side bands. For instance, the IR side bands seen in correspondence to the $a_g\nu_3$ DMTTF mode of the complexes in Ref. 15 can be interpreted as combination modes rather than as NIDW signatures. By taking half the frequency difference between high and low frequency side bands of Fig. 6 of Ref. 15, one gets a plot similar to our Fig. 4, giving clear evidence for a far-IR soft mode. Similar analysis can be done for other compounds in Table 1 with IR side bands. Among them, DMTTF-BA and TTF-2,5QCl$_2$H$_2$ show no evidence for softening of the far-IR mode, and indeed these two compounds do not dimerize down to the lowest measured $T$. There are certainly experiments which may be better interpreted in terms of domains and NIDWs, however. Moreover, NIDWs and defect states may become particularly important for pressure-induced NITs that we have not discussed here.

The complexes with *constant* $\rho$ in Table 1 are interesting, as they are subject to the Peierls instability only. Therefore they offer opportunities to study Peierls-type transitions separately from the collective CT associated with the NIT. We remark that dimerization is attributable to a spin-Peierls mechanism only in the case of fully ionic complexes, namely in the limit $\Gamma/t \ll -1$, where only spin degrees of freedom contribute. On decreasing $\rho$, charge modulation becomes important and involves larger energies. Hence the maximum dimerization in Figs. 2 and 3 or in Peierls-Hubbard models is always around $\rho \sim 1/2$. We can then understand why the temperature of the dimerization transition increases from 50 K in TTF-BA ($\rho \sim 1.0$)[38] to 120 K in M$_2$P-TCNQF$_4$ ($\rho > 0.9$)[42] and to ~220 K in TMPD-TCNQ ($\rho \sim 0.9$)[44], while TMPD-CA ($\rho \sim 0.6$)[43] already shows dimerized vibrational spectra at room temperature. These early systems have received little attention recently. The 300 K structures of[60] TMPD-TCNQ and[61] TMPD-CA are regular, albeit with large thermal parameters and leaving aside possible ordering of methyl groups, but no X-ray data at low $T$ has been reported. In addition, their paramagnetic susceptibility is thermally activated, which is not consistent with an ionic regular stack. Although models related to (1) have been widely used for segregated stacks with constant ρ and variable number of sites per repeat unit, their application to CT salts with constant ρ < 1 has been limited.

The definition of $P$ for extended chains makes it possible to evaluate the IR intensity of the dimerization mode.[32] In agreement with available far-IR data,[16] there is a huge peak at $\Gamma_P$ as seen in Fig. 2 for stacks with different stiffness. This peak is also related to the large peak observed in the dielectric constant of systems undergoing a dimerization phase transition near to the charge crossover.[15,16] The static polarizability (5), $(dP/dF)_0$, contains two terms.[62] The purely electronic contribution corresponds to the $P$ derivative taken at fixed nuclei. Its calculation requires the extension of the $P$-formalism to introduce a static electric field in the Hamiltonian.[53] The second contribution accounts for the nuclear rearrangement due to the field and is related to the IR intensity.[32] Higher-order electronic responses to a static field can also be obtained for correlated insulators[53] because $P$ is a gs property.

The data on CT salts in Table 1 point to richer behavior than the neutral-regular or ionic-dimerized stacks of TTF-CA. We expect, on general grounds, that dielectric peaks mark the Peierls transition, and we are modeling these interesting systems.[53] Lattice stiffness is an important and so far empirical parameter. Polar D or A introduces energetic and structural disorder in single crystals. Energetic disorder can be treated in greater detail than done here and can be combined with $P$ to discuss disorder-induced IR intensity in systems with variable ρ(T). By contrast, CT salts with constant ρ have Peierls-type transitions without any valence change. As noted at the beginning, neutral-ionic or Peierls transitions or ionicity changes at modest temperature or pressure involve a delicate balance among electronic ground states that are almost degenerate. These rare systems are ideal for studying electron-electron correlation, electron-phonon interactions, and a host of nonlinear interactions.

**Acknowledgments.** Work in Parma was supported by the Italian Ministery of Instruction, University and Research (M.I.U.R.) and by the Inter-university Consortium of Science and Tecnology of Materials (I.N.S.T.M., project PRISMA 2002). Work at Princeton was supported in part by the National Science Foundation through the MRSEC program under DMR-9400632.